\documentclass[9pt,twocolumn,twoside]{osajnl}
\journal{ol} 
\setboolean{shortarticle}{true} 
\ifthenelse{\boolean{shortarticle}}{\colorlet{color2}{color2b}}{\colorlet{color2}{color2}} 

\title{Compressive Raman imaging with spatial frequency modulated illumination}
\author[1]{Camille Scotté}
\author[1]{Siddharth Sivankutty}
\author[2,3,4]{Patrick Stockton}
\author[2,3,4]{Randy A.~Bartels}
\author[1,*]{Hervé Rigneault}

\affil[1]{Aix Marseille Univ, CNRS, Centrale Marseille, Institut Fresnel, F-13013 Marseille, France}
\affil[2]{W.M. Keck Laboratory for Raman Imaging of Cell-to-Cell Communications, Colorado State University, Fort Collins, CO 80523}
\affil[3]{Department of Electrical and Computer Engineering, Colorado State University, Fort Collins, CO 80523}
\affil[4]{School of Biomedical Engineering, Colorado State University, Fort Collins, CO 80523}
\affil[*]{Corresponding author: herve.rigneault@fresnel.fr}
\dates{Compiled \today}

\begin{abstract}
We report a line scanning imaging modality of compressive Raman technology with spatial frequency modulated illumination using a single pixel detector. We demonstrate the imaging and classification of three different chemical species at line scan rates of 40 Hz.
\end{abstract}
\begin{document}
\maketitle
Hyperspectral Raman imaging is an effective technique to characterize molecular species, with broad applications ranging from bio-medical research to industrial quality control. Typical implementations rely on spectrally-resolved measurements of the Raman scattered light with array detectors. This measurement of a complete vibrational Raman spectrum per spatial pixel, coupled to the weak spontaneous Raman cross-section and detector array noise, leads to lengthy acquisition and limits its implementation to slow dynamics. In situations where hyperspectral measurements aim to map the spatial distribution of molecules, the spectral data is unmixed in a post-processing step to detect molecular species and/or estimate their concentrations \cite{Keshava2002,Zhao2016,Palkki2010}. In those cases, acquiring a complete vibrational spectrum per spatial pixel may be inefficient and a massive speed up can be achieved by encompassing compressive techniques in the acquisition process.\\ 
A number of reports, termed, collectively, compressive Raman technology (CRT) have averted the acquisition of entire hyperspectral stacks \cite{Wilcox2012, Wilcox2013, Buzzard2013, Refregier2018, Scotte2018, Cebeci2018}, instead designing the measurement according to the quantities of interest to be estimated (e.g. molecular concentrations). Accurately chosen spectral components, that optimize the estimation precision, are combined on to a single pixel detector. This spectral selection is performed with a fast programmable optical filter, typically a digital micromirror device (DMD), and a single pixel detector that integrates the power spectral density of the filtered Raman spectrum . This single pixel scheme speeds up the acquisition particularly due to i) the fast temporal response of single pixel detectors, ii) the fast update rate of DMDs that can exceed conventional cameras, and iii) a larger photon flux per pixel on the detection channel. These developments have resulted in impressive demonstrations in microspectroscopy with pixel dwell times of the order of 30 $\mu$s~\cite{Wilcox2012, Rehrauer2017}. For a more detailed comparison of CRT-based estimation and state-of-the-art Raman hyperspectral imaging, we refer the readers to \cite{Scotte2018}.\\ 
While hyperspectral Raman imaging can exploit both dimensions of array detectors to perform line scanning of the sample \cite{Bowden1990, Hutchings2009, Schlucker2003}, or to perform fast widefield imaging with spatial encoding \cite{Galvis-Carreno2014, Thompson2017, Choi2017}, CRT, so far, only permits a point scanning imaging modality. In this context, we take further advantage of the single pixel architecture of our system to supplement CRT with a spatial domain multiplexing, allowing line scan imaging. By exploiting spatial frequency-modulated illumination imaging techniques (SPIFI) \cite{Futia2011, Sanders1991}, we propose and demonstrate a new strategy, Compressive Raman imaging with SPatial frequencY modulated illumination (CRiSPY). The aim is to further speed up CRT by encoding spatial information of the sample along a line in the temporal signal sent to a single pixel detector. We demonstrate line scan CRT with proof-of-concept experiments for chemically specific classification of molecular species and show that the single pixel detection with a spatially modulated line illumination affords a faster imaging rate. 

We highlight the salient principles of CRiSPY which is a combination of CRT \cite{Wilcox2012, Wilcox2013, Buzzard2013, Refregier2018, Scotte2018, Cebeci2018} and SPIFI \cite{Futia2011}. In the context of this letter, we seek to generate a spatial map of the distribution of $Q$ molecular species with known Raman spectra and estimate their proportions in each of the $N$ resolved points of the illumination line. We denote $P$ the number of illumination patterns ($p = 1...P$), $Q$ is the number of pure chemical species present in the sample ($q = 1...Q$), $M$ the number of spectral filters ($m = 1...M$) and $L$ the number of wavenumbers along a Raman spectrum ($l = 1...L$). We also note $t_p$ and $x_n$ to emphasize the spatio-temporally varying illumination patterns.\\
The $N\times Q$ matrix $\mathbf{Z}$ gathers the unknown proportions to estimate. Each of its rows $\mathbf{z}(x_n)$  specifies the proportions of pure chemical species contained in the resolved point $x_n$ of the illumination line. 
The  $P\times N$ matrix $\mathbf{W}$ contains the $P$ spatially varying illumination patterns along the line of $N$ resolved points. 
The $L\times M$ matrix $\mathbf{F}$ contains the $M$ spectral projections i.e. the $M$ spectral filters $\mathbf{f_m}$.
Each row of the $Q\times L$ matrix $\mathbf{S}$ contains the known Raman spectrum of the pure $q^{th}$ chemical species.
Then, spatial (through $\mathbf{W}$) and spectral (through $\mathbf{F}$) projections lead to the detection of photons on the single pixel detector. Those measurements are gathered in a $P\times M$ matrix $\mathbf{H}$ , which each column $\boldsymbol{\eta_m}$ contains the time trace (sampled every $t_p$) recorded when applying one spectral filter $\mathbf{f_m}$ on the DMD. 
Assuming the generated signal is linear to the excitation intensity and ignoring constant terms, this translates to: 
\begin{align}
\mathbf{H} = \mathbf{W} \mathbf{Z} \mathbf{S} \mathbf{F} 
\label{eqn:globalmatrix1}
\end{align}
Then, with $\mathbf{G^T}= \mathbf{S} \mathbf{F}$, \eqref{eqn:globalmatrix1} can be written as : 
\begin{align}
\mathbf{H} = \mathbf{W} \mathbf{Z} \mathbf{G}^T 
\label{eqn:globalmatrix}
\end{align}
While $\mathbf{Z}$ can be estimated with numerous strategies, we perform the estimation in two steps in this report. First, we seek to estimate a Raman line image for each spectral filter $\mathbf{f_m}$. In a second step, we estimate the relative proportions of the $Q$ species. The set of spectral filters $\mathbf{f_m}$ is calculated with the same optimisation procedure as in \cite{Refregier2018}. These binary amplitude filters are shown to approach the precision comparable to a hyperspectral measurement \cite{Refregier2018}. For each spectral projection, a time-varying illumination pattern $\mathbf{W}$ is generated by modulating a laser beam focused to a line with an amplitude grating whose spatial frequency is linearly swept within the acquisition window. The modulated line illumination is imaged onto the object and produces a temporal modulation of the signal, $\eta_m(t_p)$, that encodes the representation of the object in the spatial frequency domain. Hence, a line image can be obtained by a Fourier transform of the temporal signal \cite{Futia2011}. The Raman intensity in each pixel $x_n$ for the spectral filter, $\mathbf{f_m}$, is estimated through a Fourier transform and is denoted as $\hat{\eta}_m(x_n)$.   
Now, the estimation of the molecular concentrations reduces to a 1D CRT problem as in \cite{Refregier2018}:
\begin{align}
\hat{\boldsymbol\eta}(x_n) &= \mathbf{G} \mathbf{z}(x_n) 
\label{eqn:CRT}
\end{align}
With $\hat{\boldsymbol\eta}(x_n)=(\hat{\eta}_1(x_n),...,\hat{\eta}_M(x_n))^T$ and $\mathbf{z}(x_n)=(z_1(x_n),...,z_Q(x_n))^T$.
If $\mathbf{G}^T\mathbf{G}$ is not singular, the proportions $z_q(x_n)$ can be estimated with a simple least square estimation \cite{Refregier2018}:
\begin{align}
\mathbf{\hat{z}}(x_n) &= \big[ \mathbf{G}^\intercal \mathbf{G} \big]^{-1} \mathbf{G}^\intercal \hat{\boldsymbol\eta}(x_n)\
\label{eqn:CRTestimation}
\end{align}

\begin{figure}[htbp]
\includegraphics[width=\linewidth]{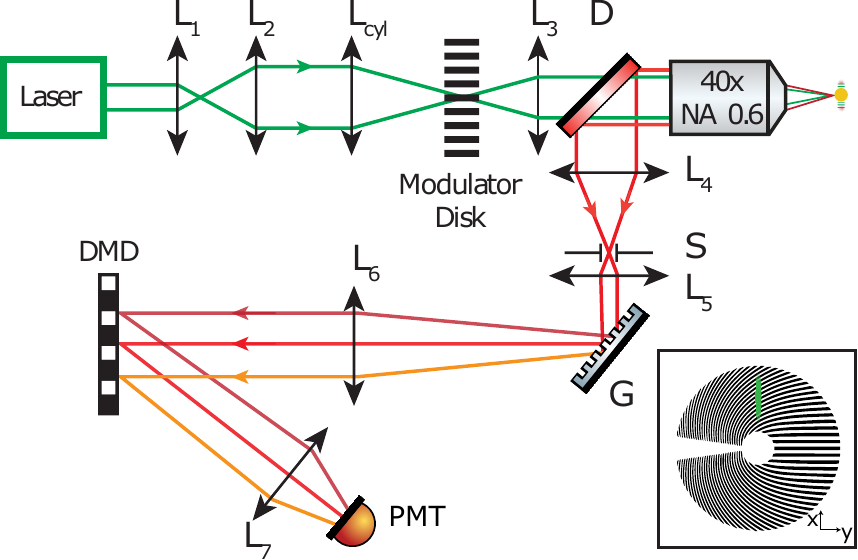}
\caption{ A schematic of the experimental setup. $L_1 - L_7$ are convex lenses with focal lengths $50$~mm, $150$~mm, $150$~mm, $100$~mm, $100$~mm, $150$~mm and $50$~mm respectively. $L_{cyl}$ - cylindrical lens with focal length $150$~mm, , D - dichroic mirror, S - confocal slit, G - amplitude grating (600~lines/mm), DMD - digital micromirror device, PMT - photomultiplier tube.}
\label{fig:Expt}
\end{figure}

\begin{figure}[t]
\includegraphics[width=\linewidth]{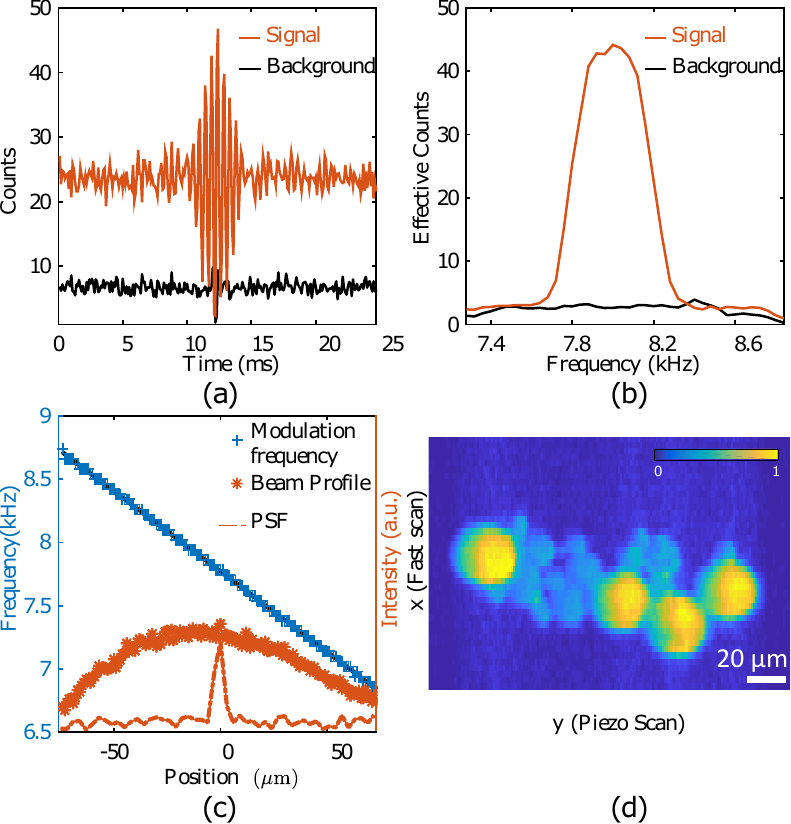}
\caption{(a) Time trace of a Raman signal of a $30~\mu$m PS bead (orange) and of the CaF$_2$ slide background (black) ($16~\mu$s binning time). (b) Corresponding isolated sidebands of the Fourier transformed time traces  (10x averaged). (c) In blue, the calibration between temporal modulation frequency and space - measured by translating an isolated $12~\mu$m MR bead. In orange, imaging system parameters - PSF and field of view (sample - MR beads). (d) Line scanned Raman image of a beads mixture (PS, PMMA, MR) on a CaF$_2$ slide with no spectral filtering.}
\label{fig:Data_demowhitelight}
\end{figure}

A simplified  schematic of the setup used in our experiments is depicted in Fig.~\ref{fig:Expt}. On the illumination side, a continuous wave laser operating at 532~\textit{nm} (Verdi, Coherent Inc) is spectrally filtered, expanded and brought to a line focus on the amplitude modulator with a cylindrical lens ($150$~\textit{mm}). This modulator is a glass disk on which an aluminium modulation pattern is imprinted (Inlight Gobos) [Inset Fig.~\ref{fig:Expt}]. This pattern takes the form (in polar coordinates)
\begin{align}
w(R,\phi) &=  \frac{1}{2} + \frac{1}{2} \mathrm{sgn} \big[ \cos(\Delta k R\phi) \big]  \ 
\label{eqn:Modulator}
\end{align}
where the sign function, $sgn$, accounts for the binary nature of the grating and $\Delta k = 10~mm^{-1}$ the finest spatial frequency on the disk. The pattern on the disk is re-imaged on to the object in a $4f-$ configuration with a de-magnification of $\approx 33$ by a combination of lenses and a microscope objective (40x, 0.6~NA). The disk finest spatial frequency, $\Delta \kappa$, and the $4f-$ imaging system limit the lateral resolution of the system to $\approx~3~\mu$m. The resolution can be improved by either increasing $\Delta \kappa$ of the modulator disc and/or increasing the NA of the imaging system. Likewise, the line length limits the field of view along $x$ to $120~\mu$m. The disk is mounted on a stepper motor (MCL-3006, Faulhaber) and is rotated at a constant velocity, $v_r$. The resulting sweep, in local spatial frequency of the modulated beam, is given by
\begin{align}
k(x,t) &= 2\pi\Delta k v_{r}t = 2\pi\kappa t \
\label{eqn:Modulator2}
\end{align} 
where $\kappa$ is the chirp parameter that relates the temporal modulation frequency to the lateral position in the object space .  
In all our experiments, $v_r = 2400$~rpm, thereby achieving line scans at a rate of 40 Hz while a piezoelectric stage scanner (P517, Physik Instrumente GmBH) holding the sample is used to scan the slow axis $y-$ yielding 2D images. On the detection side, the object plane is relayed onto a confocal slit ($100~\mu$m) to reject part of the out of focus light. A combination of dichroic mirror and notch filter ensures only the Raman signal generated from the sample is retained and is further dispersed by a blazed grating ($600~mm^{-1}$, Thorlabs) placed on the conjugate plane of the confocal slit. The spatially dispersed wavelength components of the Raman signal are imaged on to a programmable digital micro-mirror device (DMD, V-7001, Vialux -$1024\times768$ mirrors). The DMD, in conjunction with the grating, acts as a programmable spectral filter by projecting binary amplitude masks, $\mathbf{f_m}$. The effective pixel of the DMD is obtained by binning 8 mirrors along the spectral axis and all mirrors along the spatial axis to mitigate spurious diffractive effects. After binning, the spectral resolution in the current configuration is $40~cm^{-1}$, limited by the grating and the imaging system. When the DMD pixels are in the 'ON' state, under display of the filter $\mathbf{f_m}$, the corresponding spectral components of the broadband Raman signal are deflected into a photonmultiplier tube (H7421-40, Hamamatsu), while the rest is sent into a beam dump. Line images, along the $x-$ dimension, are obtained by Fourier transform of the temporal signal $\eta_m(t_p)$ recorded by the PMT and isolating its positive sidebands. Fig.~\ref{fig:Data_demowhitelight}(a) is a representative time trace of a Raman signal of a $30~\mu$m polystyrene (PS) bead (orange) and the background from the CaF$_2$ slide (black).  Fig.~\ref{fig:Data_demowhitelight}(b) shows one isolated sideband of the Fourier transformed time trace from panel (a). 

\begin{figure}[t]
\includegraphics[width=\linewidth]{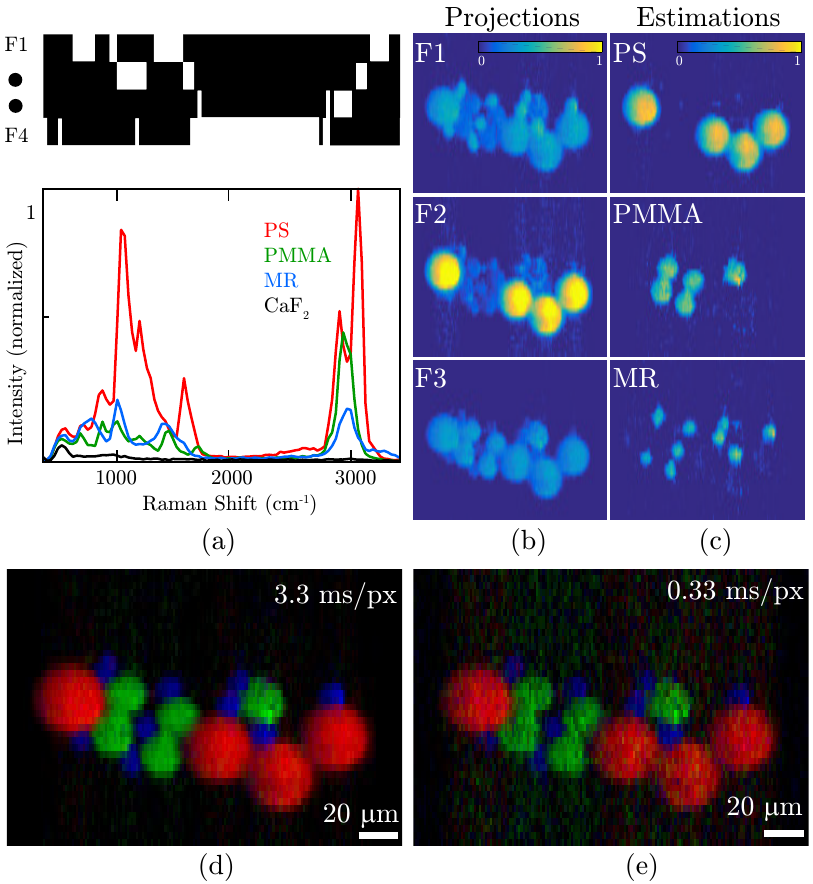}
\caption{(a) Reference Raman spectra $\mathbf{s_q}$ of the three beads and background (10 averages - 100 ms integration time per spectral bin) and representative spectral filters $\mathbf{f_m}$. (b) Images for filters $f_1$, $f_2$ and $f_3$ obtained via spatial frequency projections (normalized to the maximum intensity of the 3 images). (c) Estimated proportions of the three beads chemical concentrations (d) The composite estimated proportion maps RGB for pixel dwell time of (d) 3.3 ms and (e) 0.33 ms.}
\label{fig:Data_CS}
\end{figure}

The first step in CRiSPY imaging is to calibrate the system, i.e. measure the mapping between temporal modulation frequency and space - given by the scaling factor $\kappa$. This parameter is obtained by recording a time trace while translating an isolated $12~\mu$m melamine resin (MR) bead through the line focus and tracking the centroid of the positive sideband of the Raman signal Fourier transform as in \cite{Fields2016a}. The resulting gradient of the modulation frequency w.r.t. spatial position leads to $\kappa = 12.8$~Hz~$\mu\mathrm{m}^{-1}$ [Fig.~\ref{fig:Data_demowhitelight}(c)].
Additionally, the signal intensity from the same bead plotted w.r.t the relative $x-$ position confirms the $120~\mu$m FOV estimate. 
A spontaneous Raman image (with no spectral selectivity) of a mixture of beads (Sigma Aldrich) - $30~\mu$m polystyrene beads (PS), $20~\mu$m polymethylmethacrylate (PMMA) and  $12~\mu$m melamin resin (MR) on a CaF$_2$ coverslip (Crystran) - was obtained with spatial frequency projections [Fig.~\ref{fig:Data_demowhitelight}(d)]. Each time trace was Fourier transformed, and then the sideband corresponding to the field of view was isolated and scaled by the $\kappa$ parameter to generate a line image \cite{Field:18}. In this way, the 2D Raman image was reconstructed on a line by line basis, as opposed to point scanning.
 \\
CRT assumes the Raman spectra of the pure chemical species present in the sample are known to calculate the spectral filters $\mathbf{f_m}$. We thus measure the reference spectra $\mathbf{s_q}$ of the species present in the sample (PS, PMMA, MR and CaF$_2$ coverslip ($Q = 4$)). We improve the SNR of the calibration spectra by using the DMD as a virtual pinhole along the spatial dimension. This is a crucial step, since the line focus of our illumination patterns might also include background signal emanating from the substrate, thereby is not a chemically specific signature of the molecular species to be classified. Fig.~\ref{fig:Data_CS}(a) shows the reference spectra $\mathbf{s_q}$ (averaged on 10 spatial positions) and the associated computed filters $\mathbf{f_m}$ ($M=4$). The line images for filters $\mathbf{f_1}$, $\mathbf{f_2}$ and $\mathbf{f_3}$ obtained via spatial frequency projections are shown in Fig.~\ref{fig:Data_CS}(b). The proportion maps estimated via \eqref{eqn:CRTestimation} are shown in Fig.~\ref{fig:Data_CS}(c). The proportions were normalized and thresholded to [0 1].

\begin{figure}[htbp]
\includegraphics[width=\linewidth]{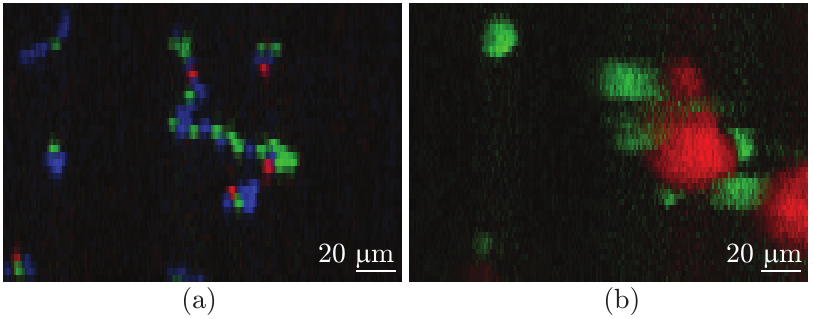}
\caption{(a) Visualization of proportion maps of PS (red), PMMA (green) and MR (blue) beads of identical sizes (5~$\mu$m) (b) Same for microcalcifications synthetic powders of HAP (green) and COM (red).}
\label{fig:Data_calcif}
\end{figure}

As a first test, we imaged a sample composed of beads with distinct chemical compositions and distinct sizes. Hence, the classification of the beads is clearly verified in the images depicted in Fig.~\ref{fig:Data_CS}(c). The total laser power along the line in this experiment was about 65 mW (irradiation $= 0.24 \times 10^{-3}  ~\mathrm{W}/\mu m^2$ ). Fig.~\ref{fig:Data_CS}(d) shows the beads distribution obtained by averaging 10 recorded time traces (corresponding to 10 sweeps of all the spatial frequencies from the SPIFI modulator), resulting in an effective dwell time $3.3~$ms per pixel - \textit{nScans$/($DiscRate} x \textit{N} x \textit{fill factor}$)$. The panel \ref{fig:Data_CS}(e) was obtained without averaging (a single sweep of all spatial frequencies) corresponding to a dwell time of $0.33~$ ms denoting the potential for rapid imaging. The signal-to-background ratio (SBR) on the PS beads is $\approx 19$ ~dB with 10 averages and $12$~dB with no averaging. We note the high fidelity classification of the four different chemical species in the case of spatially indistinguishable beads (having the same size) in Fig.~\ref{fig:Data_calcif}(a), with no further calibration (i.e. using the same spectral filters as in Fig.~\ref{fig:Data_CS}). We also demonstrate the application of CRiSPY by imaging two biologically relevant molecular species in Fig.~\ref{fig:Data_calcif}(b). Those synthetic powders mimick hydroxyapatite (HAP - green) and monohydrate calcium oxalate (COM - red) microcalcifications that may be present in the human breast and whose chemical composition relates to cancer development \cite{Kerssens2010}. The laser irradiation for the experiments in Fig.~\ref{fig:Data_calcif}(a,b) is $0.48 \times 10^{-3}  ~\mathrm{W}/\mu m^2$.

While a complete discussion of the sensitivity and limits of detection of CRiSPY is beyond the scope of this letter and dedicated to future work, we allude to certain aspects of SNR in CRiSPY. In our proof-of-concept with line illumination, all the pixels along a line are probed in parallel. Hence, the effective integration time per pixel is larger than that of a point scanner with the same line scanning rate. For an illumination power density identical to point scanning, this translates to an increase \cite{Berto2017} of about $N/2$ (if about half of the line pixels are illuminated at any instant) of the number of detected photons per bin time $\eta_m(t_p)$ - compared to point scanning. As spontaneous Raman imaging is typically photon starved, this is beneficial when operating with non-optimal detectors with additive noise. However, in our current demonstration, our measurements are shot noise limited \cite{Scotte2018}, so the noise scales with the square-root of the number of photons arising from the $N/2$ illuminated pixels across the line. While the average number of photons at pixel $x_n$ is equivalent to the power spectral density of the trace evaluated at its corresponding modulation frequency, the signal from pixel $x_n$ is affected by photon noise coming from the average Raman signal along the entire line. \\
Moreover in realistic conditions, different molecules exhibit inhomogeneity in brightness due to their distinct Raman scattering cross-sections. This results in varying SNR through the image. This disadvantage is partially mitigated in CRiSPY since each spectral filter $\mathbf{f_m}$ selects the resonances of the $q$th species. Hence, we surmise that CRiSPY would bring a SNR advantage over point scanning CRT when samples are relatively sparse \cite{diebold2013, Bialkowski1998,Fuhrmann2004, Studer2012} or when the experimental configuration exhibits signal-independent noise contributions. Further improvement could potentially be obtained by replacing the Fourier transform estimation with noise-designed estimation strategies \cite{Bialkowski1998,Fuhrmann2004}. 
In the presence of turbid medium, the ballistic components of the beams will still contribute to illumination pattern, but the CRiSPY image would degrade when the scattering is significant. Similar single pixel imaging was demonstrated on biological samples \cite{Studer2012} and through scattering media \cite{Duran2015}. SPIFI robustness to scattering in the collection path was shown in \cite{Higley2012}.\\ 
To summarize, we have reported a novel line scanning modality of compressive Raman imaging with a single pixel detector. A 40 \textit{Hz} line scanning rate is demonstrated with a rotating disk modulator offering a cost-effective speedup to conventional CRT based imaging. This opens the avenue  for Raman imaging of dynamic processes in bio-chemical systems.

\paragraph{Funding Information} We acknowledge financial support from the Centre National de la Recherche Scientifique (CNRS), Institut Carnot STAR (IRAC 2018), Aix-Marseille University A$^\ast$Midex (noANR-11-IDEX-0001-02 and A-M-AAP-ID-17-13-170228-15.22-RIGNEAULT), ANR grants France Bio Imaging (ANR-10-INSB-04-01) and France Life Imaging (ANR-11-INSB-0006) infrastructure networks and Plan cancer INSERM PC201508 and 18CP128-00. C.S. has received funding from the European Union’s Horizon 2020 research and innovation program under the Marie Skłodowska-Curie grant agreement No713750. 
\paragraph{Acknowledgements} The authors ackowledge fruitful discussion with P. Réfrégier and F. Galland.

\bigskip
\bibliography{CRISPY}

\ifthenelse{\equal{\journalref}{ol}}{%
\clearpage
\bibliographyfullrefs{CRISPY}
}{}

\ifthenelse{\equal{\journalref}{aop}}{%
\section*{Author Biographies}
\begingroup
\setlength\intextsep{0pt}
\begin{minipage}[t][6.3cm][t]{1.0\textwidth} 
  \begin{wrapfigure}{L}{0.25\textwidth}
    \includegraphics[width=0.25\textwidth]{john_smith.eps}
  \end{wrapfigure}
  \noindent
  {\bfseries John Smith} received his BSc (Mathematics) in 2000 from The University of Maryland. His research interests include lasers and optics.
\end{minipage}
\begin{minipage}{1.0\textwidth}
  \begin{wrapfigure}{L}{0.25\textwidth}
    \includegraphics[width=0.25\textwidth]{alice_smith.eps}
  \end{wrapfigure}
  \noindent
  {\bfseries Alice Smith} also received her BSc (Mathematics) in 2000 from The University of Maryland. Her research interests also include lasers and optics.
\end{minipage}
\endgroup
}{}
\end{document}